\documentclass[fleqn,usenatbib]{mnras}

\usepackage{newtxtext,newtxmath}
\usepackage[T1]{fontenc}

\DeclareRobustCommand{\VAN}[3]{#2}
\let\VANthebibliography\thebibliography
\def\thebibliography{\DeclareRobustCommand{\VAN}[3]{##3}\VANthebibliography}

\usepackage{graphicx}	% Including figure files
\usepackage{amsmath}	% Advanced maths commands

\title[MOG as symmetry breaking]{MOG as symmetry breaking in Scalar-Vector-Tensor gravity}

\author[S. Rouhani, S. Rahvar]{
Shahin Rouhani,$^{1}$\thanks{srouhani@sharif.edu}
Sohrab Rahvar $^{1}$\thanks{rahvar@sharif.edu}
\\
$^{1}$
Physics Department, Sharif University of Technology, Tehran 11365-9161, Iran}

\date{Accepted XXX. Received YYY; in original form ZZZ}

\begin{document}
\label{firstpage}
\pagerange{\pageref{firstpage}--\pageref{lastpage}}
\maketitle

\begin{abstract}
The Modified Gravity Model (MOG) has been proposed as a solution to the dark matter problem, but it does not meet the gauge invariant condition. The aim of this work is to propose a gauge-invariant theory, which suggests that symmetry can break at a low temperature in the Universe, leading to the MOG theory. This theory has the potential to alter the dynamics of the early and late Universe and naturally produce cosmological inflation.
\end{abstract}


%%%%%%%%%%%%%%%%%%%%%%%%%%%%%%%%%%%%%%%%%%%%%%%%%%

%%%%%%%%%%%%%%%%% BODY OF PAPER %%%%%%%%%%%%%%%%%%

\section{Introduction}
\label{introduction}

According to observations of the dynamics of galaxies, the cluster of galaxies, and the Universe, a significant portion of the Universe's mass is missing. This missing mass is now called dark matter \citep{silk,dark2}. Various candidates have been proposed for dark matter, including Weakly Interacting Massive Particles (WIMPs) \citep{wimp}, Axions \citep{axion}, Sterile Neutrinos \citep{sn}, and Gravitinos \citep{rioto}, as well as macro structures such as Primordial Black Holes (PBHs) \citep{car,rahvar1}, Massive Compact Halo Objects (MACHOs) \citep{Alcock,eros,rahal} and Cosmic Strings \citep{kible}.

The detection of MACHOs through gravitational microlensing in the Galactic halo in the direction of Large and Small Magellanic Clouds, has ruled out MACHOs as a dark matter candidate \citep{moniez}. Also, PBHs were limited to two specific windows of mass ranges as a potential dark matter candidate from the results of several observations \citep{carr,Kalantari_2021}. Also, experiments in particle physics, including LUX \citep{lux}, XENON100 \citep{Xenon}, SuperCDMS\citep{supercdm}, PandaX\citep{panda}, DEAP \citep{deap}, CRESST\citep{crest}, DAMIC\citep{damic}, and CoGeNT \citep{cogent}, have failed to detect dark matter particles. Additionally, space-based X-ray telescopes like Chandra X-ray, XMM-Newton, and Fermi have not observed any excess of dark matter decay \citep{acerman,fermi}.

One possible solution to the dark matter problem is modifying the law of gravity. Milgrom attempted to modify the Newtonian dynamics and the Poisson equation, which can provide reasonable solutions for the dynamics of galaxies \citep{milgrom,hodson2017generalizing,PhysRevLett.127.161302}. However, this approach fails to explain the dynamics of galaxy clusters, large-scale structure formation, and gravitational lensing as a non-covariant theory. Another attempt is the scalar-tensor-vector theory for gravity called MOG \citep{moffat}. This is a covariant theory and can explain the dynamics of galaxies and clusters of galaxies \citep{rahvar2,rahvar3}. However, MOG has a theoretical problem as the vector field in which the theory is not gauge invariant. Our work aims to present a gauge invariant theory for MOG. Gauge invariance guarantees the conservation of mass and energy. We create a short-range vector field with mass 
through a symmetry-breaking mechanism at low temperatures. This theory has additional consequences for the dynamics of the Universe.

In section (\ref{mog}) we introduce the extension of MOG, the spontaneous symmetry breaking, and its effect on the weak field approximation. In section (\ref{early}) we discuss the effect of symmetry breaking in the early Universe and inflation as a natural consequence of this theory. The conclusion is given in section (\ref{conc}). 

%In section (\ref{mog}) we will introduce the MOG theory. In section (\ref{sym}), we introduce a gauge invariant theory and show how with spontaneous symmetry breaking, MOG can be recovered. In section (\ref{cons}), the consequences of this theory will be discussed. The conclusion will be given in (\ref{conc}). 
\section{MOG and extension of the theory}
\label{mog}
 The scalar-vector-tensor gravity, so-called MOG is the extension of Einstein's gravity by introducing the extra vector and scalar fields as the gravity fields. The respective extra forces can be tuned to explain away the galactic rotation curves and clusters without a need for dark matter. 
 %The overall action can be expressed as follows. 
 Adopting the signature convention of $(-,+,+,+)$, the total action is \citep{moffat}. 
$$S = S_g + S_\phi + S_s + S_M$$

The tensor part of this theory is similar to the Einstein-Hilbert action as 
\begin{equation}
\label{g}
S_g = \frac{1}{16\pi}\int\frac{1}{\zeta}(R - 2\Lambda) \sqrt{-g}d^4x
\end{equation}
where $R$ is the Ricci scalar, $\Lambda$ is the cosmological constant. The vacuum expectation value of the field $\zeta$ is taken to be $G$, which is not exactly Newton’s constant but related to it (see below). The zeta field is a vehicle to connect the strength of the gravitational field to the expectation value of the order parameter hence the phases of the theory. 

The vector action is defined as: 
\begin{equation}
\label{vec}
S_\phi = \int(-\frac{1}{4} B^{\mu\nu}B_{\mu\nu} - \kappa A_\mu J^\mu)\sqrt{-g}d^4x, 
\end{equation}
where $B_{\mu\nu} = \partial_\mu A_\nu - \partial_\nu A_\mu $ similar to the electromagnetic field strength. %J^\mu=(\rho,J^i)$ is the current of matter and $\omega$ is the coupling constant between the matter and the vector field. 
Next comes the scalar field. In \cite{moffat} what was taken was not gauge invariant. Here we propose a gauge invariant version of \cite{moffat} which therefore involves a complex field $\varphi$ such that $\varphi\bar{\varphi} = \zeta$. This is the major difference with \cite{moffat}. So the action is:
\begin{equation}
\label{phisfield}
    S_\varphi = \int\left(-\overline{D^\mu\varphi}D_\mu\varphi -V(\varphi\bar{\varphi})\right)\sqrt{-g} d^4x
\end{equation}
where $D_\mu\varphi$ is the covariant derivative with respect to $A_\mu$: 
\begin{equation}
D_\mu\varphi = \partial_\mu\varphi + i\omega A_\mu\varphi.    
\end{equation}
A gauge transformation on the field $\varphi$ leaves this action invariant. The charge associated with the vector field is $\omega$ and the potential is taken as :
%\begin{equation}
 %  V(\varphi\bar{\varphi}) = t %\bar{\varphi}\varphi + %b(\varphi\bar{\varphi})^2
%\end{equation}
\begin{equation}
V(\varphi\bar{\varphi}) = m^2 (\varphi\bar{\varphi}) + \theta(\varphi\bar{\varphi})^2 + \frac{b}{3} (\varphi\bar{\varphi})^3
\end{equation}
where $b>0$, $\theta=  (T-T_c)/T_c$  is the reduced temperature. This potential mimics the Coleman-Weinberg mechanism i.e. a first-order transition can happen in the early universe \citep{colman,oldinf}. However, Coleman-Weinberg took a radiative corrected potential which essentially produces the same dynamics.

\subsection{Spontaneous symmetry breaking}
\label{pt}
\begin{figure}
\center
\includegraphics[width=0.5\textwidth]{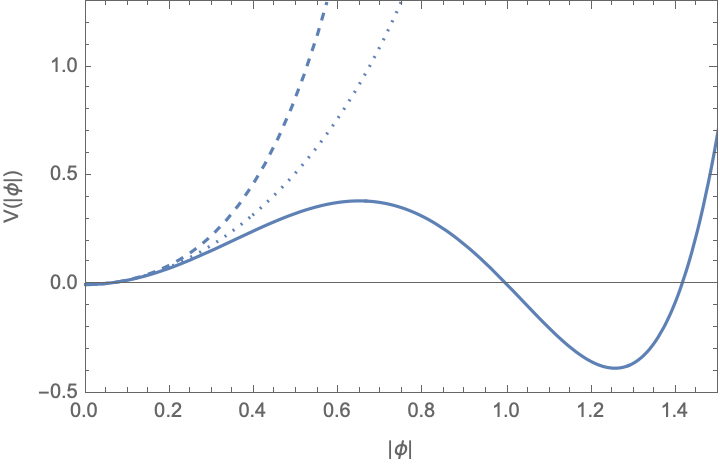}
\caption{\label{fig1} Schematic potential of meta stable $\varphi$-field for $T>T_c$ with the parameters of $m^2 = 2$, $\theta=5$ and $b = 1/3$ (dashed-line), for $T=T_c$ dotted line and the potential for $T<T_c$ with the parameters of $m^2 = 2$, $\theta=-3$ and $b = 1/3$ (solid line).}
\end{figure}

Assuming the starting temperature is near the Planck temperature, the vacuum expectation value of the scalar field vanishes, in other words, the gravitational constant vanishes i.e. we have $R = 2\Lambda$.  This temperature is much larger than GUT temperature. As $T$ reduces, at some point the reduced temperature (i.e. $\theta$) changes sign. Still no significant phenomena’s observable as the potential has not developed a new critical point. Figure (\ref{fig1}) represents the potential $V(\varphi\bar{\varphi})$ in two different phases. For the dashed curve with $T>T_c$, $<\varphi\bar{\varphi}> = 0$ which results in the effective gravitational constant vanishes, i.e. $G=0$, dotted line for $T=T_c$ and solid line for $T<T_c$ and we would expect an effective gravitational constant.

%and $V_0$ is a constant value. Hence for subcritical temperatures symmetry breaking happens and $\varphi\bar{\varphi}$ acquires a vacuum expectation value:
%Let us assume two possible signs for $b$:
%(i) For subcritical temperatures and $b>0$ symmetry breaking happens and $\varphi\bar{\varphi}$ acquires a vacuum expectation value:
%\begin{equation}
 %   <\varphi\bar{\varphi}> = -\frac{\theta}{2b}.
%\end{equation}
%To arrive back at the conventional gravity, we need to set 
%\begin{equation}
%\label{G}
 %    -\frac{\theta}{2b} = G,~~~~   T<T_c 
%\end{equation}
%(ii) for $b<0$, $<\varphi\bar{\varphi}> =0$ is a 
For $T<T_c$, the metastable phase of potential (the solid line in Figure \ref{fig1}) decays to a lower minimum and we encounter a first-order transition which releases an amount of energy into a bubble of the new vacuum \citep{colman}. The expectation value for the field is  
\begin{equation}
    <\varphi\bar{\varphi}> = \frac{-\theta+\sqrt{\theta^2 - m^2 b}}{b}.
\end{equation}
We will discuss the cosmological consequences of this phase transition later. 
%The initial symmetric solution can be taken for $T>T_{GUT}$ at $10^{14}$ GeV. For $T<T_{GUT}$ a new minimum appears. We will discuss the cosmological consequences of this theory later. %After the symmetry breaking, the scalar field gets the value of $<\varphi\bar{\varphi}>= G$. 

Let us assume a fluctuation of the scalar field around the vacuum as $\varphi = \sqrt{G} + \phi$ for $T<T_c$ where $<\varphi\bar{\varphi}> =G$. Then
the covariant derivative around the vacuum is
\begin{equation}
D_\mu \varphi = \partial_\mu\phi + i\omega A_\mu (\sqrt{G}+\phi).    
\end{equation}
After substituting in the action, a mass term for the vector field $A_\nu$ arises. This leads to a mass term for the field $A_\nu$ as 
%\begin{equation}
   $ -\frac{1}{2}\omega^2 G A_\nu A^\nu$. 
%\end{equation}
So the overall action for $A_\nu$ can be written as 
\begin{equation}
\label{vec}
S_\phi = \int(-\frac{1}{4} B^{\mu\nu}B_{\mu\nu} -\frac{1}{2}\mu^2 A_\nu A^\nu - \kappa J^\mu A_\mu)\sqrt{-g}d^4x, 
\end{equation}
where $\omega^2 G = \mu^2$, and $J^\mu$ is the four current of matter. The charge of the $A_\nu$ field is related to the inertial mass as $q = \kappa m$. \\
Gauge invariance ensures the conservation of matter current which is no longer guaranteed by the Bianchi identities in MOG. Furthermore, a unique answer is obtained by gauge fixing.

%Here we add the mass term of $${\cal L}_{mass} = - \frac{1}{2}\mu^2\phi^\mu\phi_\mu$$ to the action. 

%The action for the scalar field is 
%\begin{equation}
%\label{eqG}
 %   S_G = \int\frac{1}{G}\left(-\frac{\nabla_\mu G \nabla^\mu G}{G^2} - V_G(G)\right)\sqrt{-g}d^4x
%\end{equation}
\subsection{Weak approximation and the effect of mass term}
In the weak field approximation \citep{moffat_rahvar}, we can perturb the fields around the background as $g_{\mu\nu} = \eta_{\mu\nu} + h_{\mu\nu}$ and take $A^\mu$ as the first order perturbation . By varying the action and keeping first-order perturbation terms, the field equation obtain as 
\begin{eqnarray}
\label{poisson}
    \nabla^2h_{00} &=& 8\pi G\rho, \\
     \nabla^2 A^0 - \mu^2 A^0 &=& \kappa  \rho,
     \label{yuku}
\end{eqnarray}
%where the contribution of the perturbation around $G_0$ from the differential equation for $G$ is a second-order term and we keep $G$ as a constant parameter. 

Equations (\ref{poisson}) and (\ref{yuku}) have analytical solutions as 
\begin{equation}
    h_{00} = -2G\int \frac{\rho(x')}{|x-x'|}d^3x'
\label{h0}
\end{equation}
and 
\begin{equation}
    A^0 = -\kappa \int\frac{e^{-\mu|x-x'|}}{|x-x'|}\rho(x')d^3x'.
\label{phi0}
\end{equation}
Now, we use the action of a point-like particle interacting with the vector field and living in the Riemannian manifold with the metric of $g_{\mu\nu}$ as 
\begin{equation}
S = -m\int d\tau +  \kappa m\int A_\mu dx^\mu,
\end{equation}
%where the charge of the particle associated to the vector field is replaced with the inertial mass as $q = \kappa m$. 
From this action, we can derive the Hamiltonian of the particle where it is also applicable for $m=0$ \citep{rahvar2}. The equation of motion of this particle is 
\begin{equation}
    \ddot{x}^\mu +\Gamma^\mu{}_{\alpha\beta}\dot{x}^\alpha\dot{x}^\beta = \kappa B^\mu{}_\nu \dot{x}^\nu.
\end{equation}
The right-hand side of this equation is similar to the Lorentz force in electromagnetism, However, in this theory, we only have positive charges, which means that the impact of this term would be repulsive. 

For the weak field approximation, the spatial component can be written as $\ddot{x}^i -\frac{1}{2} \partial^i h_{00} = -\kappa \partial^i A^0$ where the effective potential is 
\begin{equation}
\label{phi}
\Phi_{eff} = \frac{1}{2}h_{00} - \kappa A^{0}.
\end{equation}
Now we substitute the solutions from the equations (\ref{h0}) and (\ref{phi0}) in the definition of the effective potential, the result is 
\begin{equation}
    \Phi_{eff} = \int (-G + \kappa^2 e^{-\mu|x-x'|})\frac{\rho(x')}{|x-x'|}d^3x'.
\label{effp}
\end{equation}

For the large distances where $|x-x'|\gg \mu^{-1}$, we recover $1/r$ potential with the gravitational constant of $G$. Also for the close distance when $x\rightarrow x'$, the effective gravitational constant would be $(G -\kappa^2)$ which is identical to Newtonian gravitational constant (i.e. $G_N$). So we can rewrite equation (\ref{effp}) as follows:
\begin{equation}
    \Phi_{eff} = -G_N\int\left(\alpha + \left(1-\alpha\right)e^{-\mu|x-x'|}\right)\frac{\rho(x')}{|x-x'|}d^3x',
\end{equation}
where $\alpha = G/G_N$. By comparing the acceleration from the effective potential with the dynamics of spiral galaxies as well as the cluster of galaxies the free parameters of the theory obtained as $\alpha = 8.89\pm 0.34$ and $\mu = 0.042\pm0.004 \text{kpc}^{-1}$  \citep{rahvar1,rahvar3,rahvar4}.

\section{MOG in early Universe}
\label{early}
This theory has also implications for the early Universe. 
%We consider two distinct cases of (i) $T_c$ is low and happening by means of low-temperature Universe and (ii) $T_c$ is high and happening in the very early Universe. 
For $T>T_c$ (i.e. $\theta>1$), the potential behaves as shown in the dashed curve of Figure (\ref{fig1}), resulting in $<\varphi\bar{\varphi}> = 0$, or $G=0$. In this case, the gravitational field is not switched on yet, and from the action (i.e. $R = 2\Lambda$) the Universe undergoes a slow expansion due to the cosmological constant. We can have also fast inflation if we let the potential of $V(\varphi\bar{\varphi})$ has an offset from in x-axis. The Universe under the metastable state eventually undergoes the quantum tunneling mechanism from the false vacuum to the true vacuum, similar to the old inflationary scenario \citep{oldinf}. 
The main difference is that effective gravitational constant emerges from the non-zero $\varphi$ -field to the value of 
\begin{equation}
\label{G}
G = <\varphi\bar{\varphi}> = \frac{-\theta+\sqrt{\theta^2-m^2 b}}{b}.
\end{equation}
The tunneling happens through the barrier of the potential. After the tunneling to the true vacuum, the energy of potential reheats the Universe. From this point onwards, the standard model of particles applies and all standard particles emerge. Hence from this moment the Universe evolves just like the standard model of cosmology, except that the gravitational constant will be a function of the temperature with its own consequences.

%The result would be similar to the old-inflationary scenario, except that area of the universe in the false vacuum state has no gravity. 

%In the {\it second scenario}, we take $T<T_c$ where the potential profile is shown with the solid line in Figure (\ref{fig1}). This potential is similar to the slow-rolling inflationary scenario \cite{Starobinsky,guth}. In this case, the $\varphi$ field plays the role of inflaton field with $T^{\mu\nu} = -g^{\mu\nu} V(\varphi\bar{\varphi})$. The quantum fluctuations of this field can generate the seeds for the large-scale structure of the Universe.  We plot the evolution of the gravitational constant for both cases of negative and positive $\theta$ in Figure (\ref{fig2}). Here the expansion of the Universe is along with decreasing the $t$ parameter.  %temperature (i.e. here $t$ parameter) the effective gravitational constant is getting larger. 
%For $\theta\rightarrow -1$, the temperature of the Universe gets closer to zero which results in $G =\frac{1}{3\nu}(-b+\sqrt{b^2 + 3\nu})$.

\begin{figure}[b]
\center
\includegraphics[width=0.5\textwidth]{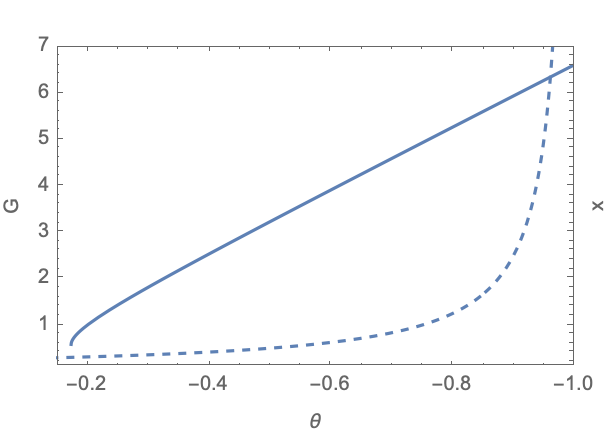}
\caption{\label{fig2} The evolution of gravitational constant, $G$ (solid line with an arbitrary scaling) and $x = a/a_c$ (dashed line)   as a function of $\theta = (T -T_c)/T_c$ (at x-axis) after the end of inflation during the radiation-dominated epoch. The decrease of $\theta$ is in the direction of the increase of time.}
\end{figure}

The dynamics of gravitational constant grow according to the equation (\ref{G}) as in Figure (\ref{fig2}) (solid line). The dependence of the $G$ on the temperature of the Universe can modify the expansion history of the Universe. Assuming the radiation-dominant epoch of the Universe, the scale factor depends on temperature as $a\propto 1/T$. So the kinematics of the Universe in terms of $\theta$ parameter is $x = (1+\theta)^{-1}$ where $x = a/a_c$, as shown in Figure (\ref{fig2}). 
 In this theory as the Universe expands the gravitational constant grows (as shown in Figure 2). This is in contrast to the Dirac large numbers hypothesis.

%From equation (\ref{G}) to have a real value for $G$, the parameter of $x$ ranges as 
%$$  (1+m\sqrt{3b})^{-1}<x<(1-m\sqrt{3b})^{-1}.$$
%\textcolor{red}{Equation (19) has a fundamental problem. Here we have an lower limit on the temperature of the Universe. Means that }

%where $x=1$ represents $\theta = 0$ at $T=T_c$.  
%effective FRW equation can be written as 
%temperature is a free parameter and can be fixed by studying the dynamics of gravitational constant in terms of the scale factor.   

\begin{figure}
\center
\includegraphics[width=0.5\textwidth]{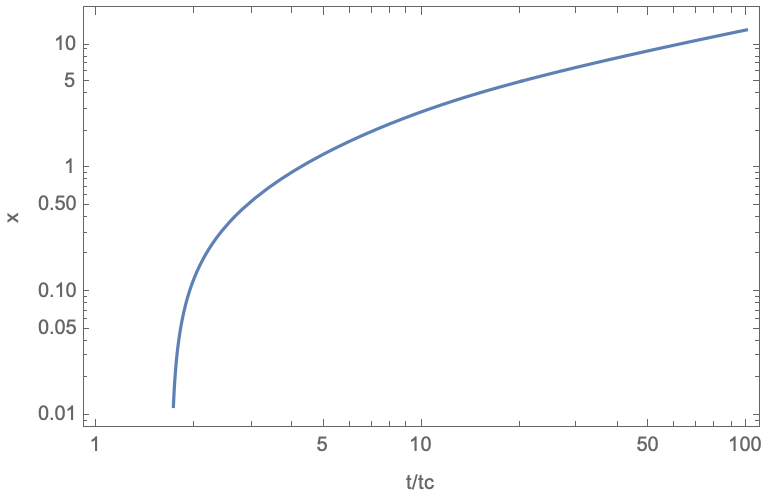}
\caption{\label{fig3} The numerical solution of equation (\ref{x}), the scale factor in terms of time in logarithmic scale. For the early Universe, the rate of expansion due to inflation is high and the effective gravitational constant is small, later $G$ grows, and the scale factor grows as $t^{1/2}$.}
\end{figure}

We can also investigate the dynamics of the Universe, as a function of time taking into account an effective dynamical $G$. From the action, the field equation for the scale factor is given by 
\begin{equation}
    (\frac{\dot{a}}{a})^2 = \Lambda + \frac{8\pi G(a)}{3} \rho_r(a), 
\end{equation}
where the matter content of the Universe is radiation and the effective gravitational constant is given by the equation (\ref{G}). Using the definition of $x = a/a_c$, the modified FRW can be written as 
\begin{equation}
x'^2 x^2 = \bar{\Lambda}x^4 + \frac{1-x^{-1} + \sqrt{(1-x^{-1})^2 - 3m^2b}}{1+\sqrt{1- 3m^2b}},
\label{x}
\end{equation}
where $' \equiv {d}/{d\bar{t}}$ and $\bar{t} = t/t_c$ and $\bar{\Lambda} = \Lambda t_c^2$. The solution of this equation is shown in Figure (\ref{fig3}) where around $t\sim t_c$ Universe undergoes a fast expansion as the Universe has a large momentum for expansion due to inflation and the effective gravitational constant is zero.  For $t\gg t_c$, the gravitational constant switches on, and the Universe enters the radiation dominant epoch with $a \sim t^{1/2}$ expansion rate. This modification to the dynamics of the Universe at later times may also shed light on the cosmic tension on $H_0$ and $\sigma_8$ \citep{riess}.

%However, the vacuum energy of the scalar field would be $V = V_0$ or equivalently $T^{\mu\nu} = V_0 g^{\mu\nu}$.   

%This condition implies that  $R-V_0 = 0$ \textcolor{red}{(Here we need $\xi$ in the dominator of equation (3))} causing gravity the Universe undergoes an inflationary area similar to the old inflationary scenario \cite{oldinf}.

%to switch off in the early Universe and the dynamics to be governed by the cosmological constant. 

%which means that in the early Universe, the gravity is switched off and the dynamics of the Universe is given by the cosmological constant.
We have also the vector field of $A^\mu$ which controls the repulsive force in the weak field approximation in equation (\ref{phi}) and it is important to determine the dynamics of this field during the expansion of the Universe. A detailed dynamics of electromagnetic fields in FRW metric can be found in \citep{mashhoon} and the vector field introduced here, will follow the same dynamics. 
%Using the FRW metric, the dynamics of the scalar factor $a$ can be described by the equation
% \begin{equation}
 %   \frac{\ddot {a}}{a} + (\frac{\dot{a}}{a})^2 = \frac{\Lambda}{3}
%\end{equation}
%with the solution $a\propto \exp(\sqrt{\frac{\Lambda}{6}}t)$. 
%The numerical value of the cosmological constant is approximately 
%$\Lambda\simeq H_0^2$, indicating that before the critical temperature, the Universe undergoes exponential expansion with a small rate of $H_0$.  
After the end of inflation, as a result of reheating particles as well as the particles associated with the vector field generated. So we consider the evolution of the vector field in the radiation-dominant epoch. 
%During the inflation, we have de-sitter phase of exponential expansion while the temperature of the Universe with the content of radiation  decreases as $T\propto 1/a$,  until the spontaneous symmetry breaking happens.  
The vector field follows the equations of $\nabla_\nu\nabla^\nu A^\mu = R^\mu{}_{\nu}A^\nu$ where we take Lorentz gauge for the vector field. The solution of this equation, taking the spatial part of the wave equation as $\exp(\pm i k.x)$, in the conformal FRW metric is 
\begin{equation}
\frac{d^2A^\mu}{d\tau^2}+\frac{4}{\tau}\frac{d A^\mu}{d\tau}+ (k^2 - \frac{3}{\tau^2})A^\mu = 0.
\end{equation}
The solution of this differential equation is the Bessel function as 
\begin{equation}
    A^\mu(\tau) = \frac{1}{\tau^{3/2}}\left(C_1 J_{\frac{\sqrt{21}}{2}}(k \tau) + C_2 Y_{\frac{\sqrt{21}}{2}}(k \tau)\right)
\end{equation}
where we use the physical time and also replace the scale factor with time as $a\sim t^{1/2}$, 
\begin{equation}
    A^\mu(x,t) = \frac{e^{\pm ikx}}{a^{3/2}}\left(C_1 J_{\frac{\sqrt{21}}{2}}(\frac{k}{a}t) + C_2 Y_{\frac{\sqrt{21}}{2}}(\frac{k}{a}t)\right).
\end{equation}
Here, the amplitude of the vector field decays as $1/a^{3/2}$ and for $(k =\omega$, letting $c=1$), the effective frequency of this field also decreases with ${\tilde \omega}= \omega/a$, as we expected from the cosmic expansion.

%space is an exponential decrease in the amplitude of the vector field while the field is oscillating with the frequency of inverse of the present age of Universe
%\begin{equation}
 %   A^\mu \propto \exp{(-\sqrt{\frac{3}{8}\Lambda}t)} \sin(\sqrt{\frac{\Lambda}{8}}t),
%\end{equation}
%where we can ignore the oscillation mode of this solution. 

%The perturbation theory of the fields as the seed for the large-scale structure formation needs detailed calculations and comparisons with the observations. 
\section{Conclusion}
\label{conc}
Summarizing this work, we aimed to cure the gauge invariance problem of Scalar-Vector-Tensor gravity (MOG). The mass of the vector field naturally obtains from introducing a complex scalar field where the derivative is the covariant definition with respect to the vector field. We could recover the weak field approximation of this theory that fits well with the dynamics of galaxies and clusters of galaxies without the need for dark matter.  In this way the need for dark matter in the cluster of galaxies, spiral and elliptic galaxies is resolved. Also, understanding the lensing in the bullet cluster may be cured of the dark matter problem, this however requires careful calculations of lensing which is different in the present theory. 
%Dense galaxies which are smaller than the range of the vector field will not need dark matter.

Moreover, the phase transition of the scalar field in the early Universe naturally causes the inflationary expansion of the Universe. Perturbations of the field during the inflation and growth of the perturbations, taking into account that the effective gravitational constant depends on the temperature of the Universe can cause substantial observational consequences.

\section*{Acknowledgments}
We would like to express our gratitude to F. Loran for the valuable discussions in this work. Also, we thank the anonymous referee for his/her useful comments improving this work.

%\bibliographystyle{plain}
%\bibliographystyle{unsrt}
%\bibliography{apssamp}

%%%%%%%%%%%%%%%
\section*{Data Availability}

No data is generated in this work.



\bibliographystyle{mnras}
\bibliography{example} % if your bibtex file is called example.bib

%\begin{thebibliography}{99}
%\end{thebibliography}

%%%%%%%%%%%%%%%%%%%%%%%%%%%%%%%%%%%%%%%%%%%%%%%%%%

%%%%%%%%%%%%%%%%% APPENDICES %%%%%%%%%%%%%%%%%%%%%

%\appendix

%If you want to present additional material which would interrupt the flow of the main paper, it can be placed in an Appendix which appears after the list of references.

%%%%%%%%%%%%%%%%%%%%%%%%%%%%%%%%%%%%%%%%%%%%%%%%%%

% Don't change these lines
%\bsp	% typesetting comment
%\label{lastpage}
\end{document}